\documentclass{mn2e}
\RequirePackage{graphicx}
\begin{document}
\title[VLA1623 on scales of $\sim$50--100AU]{The immediate 
environment of the Class 0 protostar VLA1623, on scales of $\sim$50--100~AU,
observed at millimetre and centimetre wavelengths}
\author[Ward-Thompson {\em et al.}]
{D. Ward-Thompson$^1$\thanks{e-mail: D.Ward-Thompson@astro.cf.ac.uk},
J. M. Kirk$^1$, J. S. Greaves$^2$, P. Andr\'e$^3$ \\
$^1$School of Physics and Astronomy, Cardiff University,
Cardiff, CF24 3AA\\
$^2$School of Physics and Astronomy, St Andrews University, North Haugh,
St Andrews, Fife, KY16 9SS\\
$^3$Laboratoire AIM, DSM/IRFU/Service d'Astrophysique, CEA Saclay,   
91191 Gif-sur-Yvette, France}

\maketitle

\begin{abstract}
We present high angular resolution observations, taken with the Very Large 
Array (VLA) and Multiple Element Radio Linked Interferometer Network 
(MERLIN) radio telescopes, at 7~mm and 4.4~cm respectively,
of the prototype Class 0 protostar VLA1623. At 7~mm we detect
two sources (VLA1623A \& B) coincident with the two previously detected
components at the centre of this system.
The separation between the two is 1.2~arcsec, or $\sim$170~AU at an
assumed distance of 139~pc.
The upper limit to the size of the source coincident with
each component of VLA1623 is $\sim$0.7~arcsec, in agreement with
previous findings. This corresponds to a diameter of $\sim$100~AU
at an assumed distance of 139~pc. 
Both components show the same general trend in their broadband 
continuum spectra, of a steeper dust continuum spectrum shortward
of 7~mm and a flatter spectrum longward of this. 

We estimate an upper limit to the VLA1623A disc mass of $\leq$0.13~M$_\odot$
and an upper limit to its radius of $\sim$50~AU.
The longer wavelength data have a spectral index of $\alpha$$\sim$0.6$\pm$0.3.
This is too steep to be explained by optically thin
free-free emission. It is most likely due to optically thick
free-free emission. Alternatively, we speculate that it might be due to
the formation of larger grains or planetesimals in the circumstellar disc.
We estimate the mass of VLA1623B to be $\leq$0.15~M$_\odot$.
We can place a lower limit to its size of $\sim$30 $\times$ 7 AU,
and an upper limit to its diameter of $\sim$100~AU.
The longer wavelength data of VLA1623B also have a spectral index of 
$\alpha$$\sim$0.6$\pm$0.3.
The nature of VLA1623B remains a matter of debate.
It could be a binary companion to the protostar, or a knot in the radio jet
from VLA1623A.
\end{abstract}


\section{Introduction}
\label{intro}

Young stellar objects (YSOs) are divided observationally into a number of 
different classes. Class 0 protostars represent the earliest observed 
phase of protostellar evolution (Andr\'e, Ward-Thompson \& Barsony, 1993; 
hereafter AWB93), in which the protostar has accumulated less than half
of its final main-sequence mass. The subsequent Class I phase (Lada 1987;
Wilking, Lada \& Young 1989) occurs when the protostar accumulates the
majority of the remainder of its mass. By the time of the Class II
phase (Lada 1987; Wilking et al., 1989), the large-scale envelope
seems to have entirely
accreted onto the protostar or its circumstellar disc
(e.g. Ward-Thompson, Andr\'e \& Lay 2004; Ward-Thompson 2007). This phase
generally correlates with
the `classical' T~Tauri stage (Andr\'e \& Montmerle 1994).

There remain some aspects of protostellar evolution that are 
still a matter of debate. For example, the epoch of
circumstellar disc formation is not precisely
known, and the time of the onset of planet formation within the
disc is also highly uncertain. 
There are as yet insufficient observations
to tie down the beginnings of these important phases.
Furthermore, the effect of binarity on this simple picture is not
fully understood.

Indirect evidence for the presence of a disc is
usually implied from the presence of
a bipolar jet or outflow (e.g. Curiel et al. 2006; Torrelles et al. 2011),
and jets and outflows have been observed in the
Class 0 stage (e.g. Bontemps, Ward-Thompson \& Andr\'e 1996;
Andr\'e, Ward-Thompson \& Barsony, 2000 -- hereafter AWB00).
In fact, bipolar outflows become less energetic 
and less well collimated with age 
(Henriksen, Andr\'e \& Bontemps 1997). 
Consequently, it is implied that discs must form in the earliest
protostellar stages. 
There is also evidence that
they are threaded by magnetic fields
(Holland et al. 1996; Girart et al. 2006).

Single-dish observations with current mm telescopes typically provide 
resolutions of order 1000--2000~AU in nearby star-forming 
regions, and so probe circumstellar envelopes.
Interferometers can reach scales below 100~AU, and so can probe the 
structure of circumstellar discs (e.g. Guilloteau et al. 2008;
Dutrey et al. 2008; Schaefer et al. 2009).
A comparison between interferometer data 
and single-dish data allow one to assess the relative importance of 
circumstellar discs and envelopes (Ward-Thompson 2007). For Class 0 
protostars the disc contains typically $\leq$20-25\% of the total 
mm flux. For Class I sources this value is typically around $\sim$25\%. 
For Class II sources, typically all of the mm flux arises from the disc
(Ward-Thompson et al., 2004; Ward-Thompson 2007).

VLA~16234-2417 (hereafter
VLA1623) was the first Class 0 object to be recognised, and it is still
one of the youngest known members of this class (AWB93). 
VLA1623 has a jet (Bontemps \& Andr\'e 1997), 
a bipolar CO outflow 
(Andr\'e et al.,1990; Dent et al., 1995),
and evidence for a circumstellar
disc (Pudritz et al., 1996).
This would indicate that discs, like jets and outflows, are 
formed at the very beginning of protostellar evolution.

Subsequent observations at 2.7~mm suggested that VLA1623 was in fact a 
binary protostar, and that a source which had previously been identified
as a knot in the jet was in fact a second protostellar component
(Looney, Mundy \& Welch, 2000; 2003). Somewhat confusingly, 
the second component was originally known as `knot A' of the jet
(Bontemps \& Andr\'e 1997), and was then renamed `component B'
of the protostar (Looney et al., 2000). We follow the more recent
naming system and refer to the originally known protostar as VLA1623A,
and to the second component to the west as VLA1623B, after 
Looney et al., (2000).

A number of circumstellar discs have now been imaged around young stars,
although the majority of those studied have been more evolved pre-main
sequence stars. For example, Qi et al. (2004) imaged the disc around
the Class II source TWHya, and found it to be consistent with a disc in
Keplerian rotation.

The question as to when planets begin to form in discs is also very much
a matter for debate. A low-mass companion was found in the disc of the
Class II protostar HL~Tau (Welch et al., 2004). It has recently been
hypothesised that this may be a planet in the process of formation
(Greaves et al., 2008). If so, then this indicates that planets,
or at least planetesimals, have started to form by the Class II stage.
However, subsequent work on HL~Tau did not see the companion source
(Carrasco-Gonzalez et al. 2009).

Planetesimal literally means a piece of a planet. Typical
sizes of planetesimals are kilometer-sized or larger. These are
the building blocks of the rocky planets like the Earth, and also of
the cores of gas giants such as Jupiter. 
One problem in our understanding of
the formation of planetesimals is that of drag between
the gas and the solids. 

The size of a solid particle will determine how it is affected by drag
(Brauer et al., 2008). Dust grains of a millimetre or less in
diameter will co-orbit with the gas. However, objects 
of centimetre to metre size
experience significant drag from the surrounding gas. Models predict
that such objects would rapidly lose
angular momentum and spiral into the central star. Objects significantly 
larger than this, namely planetesimals of
kilometre size or greater, have sufficient inertia that they are 
effectively immune to drag. 
Collisional fragmentation is also a potential barrier to grain growth
(Brauer et al., 2008). Therefore,
whatever process takes material from millimetre to kilometre 
sizes must occur fairly rapidly otherwise all of
the material would end up in the central star.

There are very few observations of large grains or planetesimals in discs 
to constrain these models. The peak of their emission should be shifted
to longer wavelengths relative to `normal' inter-stellar dust grains.
Hence, it is necessary to search for them at mm and cm wavelengths.
In this paper we present high angular resolution observations of VLA1623, 
that were taken in the mm and cm regimes, in order to probe the 
immediate environment of the protostar itself and study its 
disc, at this very earliest stage of protostellar evolution.

\section{Observations}
\label{obs}

We present data from the Very Large Array (VLA) radio telescope, which
consists of 27 radio antennas in a Y-shaped configuration,
located near Socorro, New Mexico, in the USA. 
Each antenna is 25~m in diameter. The antennas can be
moved from a dispersed configuration (A array) with a maximum baseline
of 36~km, to a compact configuration with a maximum baseline of 1~km
(D array). We show previously unpublished data from the VLA archive. 
The data were taken on 2004 February 1 in B/C configuration
($\sim$4--10-km baselines), and on 2005 July 29
in C array (3.6~km maximum baseline), at a wavelength of 7mm (43GHz).

\begin{figure}
\includegraphics[angle=0,width=80mm]{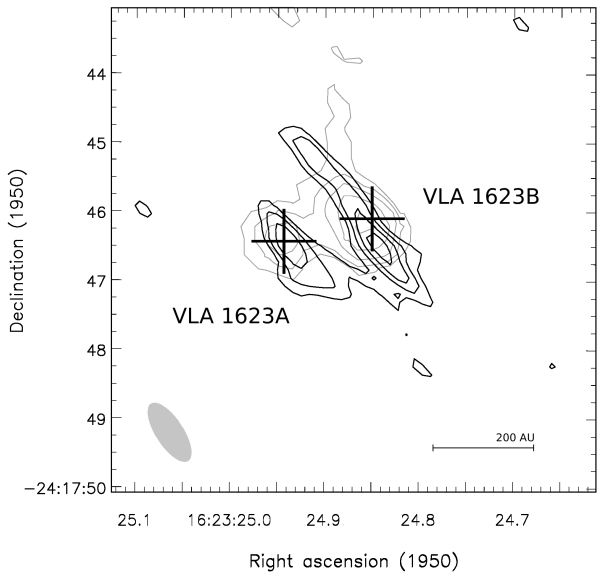}
\caption{Isophotal contour map of the new 7-mm data
of VLA1623, plotted in 1950.0 coordinates for ease of comparison with
older data. The lowest contour is at 3~$\sigma$, and the contour
interval is 1~$\sigma$. The 1-$\sigma$ level of the map is 0.2~mJy/beam.
The lighter grey overlaid contours are the 3.6-cm data of
Bontemps \& Andr\'e (1997). The position of the VLA1623A protostar 
is marked with a cross, as is the source immediately to the west, labelled 
VLA1623B (originally known as `knot A'). 
The peak positions of the two sources at 7~mm are
R.A.(1950) = 16$^{\rm h}$23$^{\rm m}$24.935$^{\rm s}$, 
Dec.(1950) = $-$24$^\circ$17$^\prime$46.5$^{\prime\prime}$ for VLA1623A, and
R.A.(1950) = 16$^{\rm h}$23$^{\rm m}$24.840$^{\rm s}$, 
Dec.(1950) = $-$24$^\circ$17$^\prime$46.5$^{\prime\prime}$ for VLA1623B.
The beam-size for the 7-mm data is 1.0 $\times$ 0.4 arcsec, at a
position angle of 33 degrees, and is shown as a grey ellipse
in the lower left-hand corner
(the beam-size for the 3.6-cm data was 
1.0 $\times$ 0.75 arcsec, at a position angle of 26 degrees).
A scale bar of 200~AU is shown at lower right.}
\label{fig1}
\end{figure}

The data were reduced in the standard fashion, using the software 
package AIPS (Bridle \& Greisen, 1994). 
Natural beam weighting was used and 500 cleaning iterations
were carried out. Phase reference 
calibration was performed using the standard reference source 
16258-25276. The flux calibrator used was 1331+305. The 
final full-width at half maximum (FWHM)
beam-size of the data was 1.0 $\times$ 0.4 arcsec, at a
position angle of 33 degrees. 
The 1-sigma rms noise on the final map was found to be 0.2~mJy/beam.

We also obtained data from the 
Multiple Element Radio Linked Interferometer Network (MERLIN)
at a wavelength of 4.4cm (6.8GHz). 
MERLIN is an array of six observing stations that together form 
a telescope with an effective aperture diameter
of over 217 kilometres, whose control centre is based
at Jodrell Bank, near Manchester, in the UK. 
Observations were taken on several days 
between 2009 January 23 and February 24. A total of 37 hours of on-source 
integration time was taken. 

The very low declination of the source 
($-$24 degrees) meant that the source was only visible to MERLIN for a few 
hours each day, and never rose higher than an elevation of 14 degrees. 
Subsequently, it was found that the data at each end of every track was of
much lower signal to noise ratio than the rest. In addition, some parts of 
some tracks also suffered from significantly lower than average signal to 
noise ratio, due to the low elevation of the source. These data were 
discarded, leaving data of total integration time $\sim$25 hours.
The standard source 1622-253 was used for phase and amplitude calibration. 
Natural beam weighting was used. The final 
FWHM beam-width of the data was found 
to be $\sim$0.2 $\times$ 0.05 arcsec, at position angle 9 degrees. 
The 1-sigma rms noise level in the final map was seen to be 
$\sim$80~$\mu$Jy/beam. 

The infra-red source 
S1 was also in our mapped area (Wilking et al., 1989). This is known to be a 
young magnetic B3 star, and is a radio source 
sometimes referred to as LFAM11 (Leous et al., 1991). 
In radio observations it is seen to have a core-halo structure, with
synchrotron emission coming from the star's magneto-sphere, and 
free-free emission from the surrounding compact HII region.
The MERLIN observations are more likely to be sensitive to the core
component. This is seen to have a flat spectrum between 2~cm and 6~cm,
with a flux density of $\sim$8~mJy throughout this range 
(Andr\'e et al., 1988). We detected LFAM11 strongly 
in our data, and recovered a 4.4-cm flux density of $\sim$6~mJy,
consistent with the earlier data.
This gave us added confidence that the data taking and 
reduction methods had all worked correctly, despite the difficulty of 
observing at this low declination, and from this we estimate that
the absolute calibration accuracy of our MERLIN data is $\sim$25--30\%.

\section{Results}
\label{res}

Figure 1 shows the 7-mm VLA map of VLA1623 as an isophotal contour map. 
Two sources are seen in the image, with the western source appearing
marginally more extended, while the eastern source is more point-like. The 
apparent extension of both sources in a north-easterly to south-westerly 
direction is simply a product of the beam shape of these data. 
The beam is shown as a grey ellipse in the lower left-hand corner.

The flux density of the western source is 2.0$\pm$0.2~mJy, 
while that of the eastern source is 1.2$\pm$0.2~mJy.
Superposed on Figure 1, in lighter grey contours, are the 3.6-cm VLA 
data of Bontemps \& Andr\'e (1997). 
These data also show two sources, with the 
western source having a 3.6-cm
flux density of 0.40$\pm$0.02~mJy, and the eastern source being 
0.13$\pm$0.02~mJy. All of these flux densities are listed in Table~1.

The crosses on Figure~1 show the positions of the two components of
the system, A \& B, identified at 2.7~mm by Looney et al., (2000).
Once again we see that these components are coincident with the two
sources we have detected at 7~mm.
Hence we see that we have detected both components, 
and resolved them from one another in the 7-mm data. 

The eastern component is agreed by all to be
the centre of the protostar VLA1623, which we shall 
refer to hereafter as VLA1623A (after Looney et al., 2000).
The nature of the second component has been a matter of debate.
Bontemps \& Andr\'e (1997) hypothesised that it was a knot of the 
jet emanating from the protostar. Looney et al., (2000)
suggested it was a second protostar, which they called VLA1623B.
We follow the latter naming system
and hereafter refer to the western source as VLA1623B.
We note that the position of VLA1623A remains constant to within
the astrometric accuracy of the different interferometers. However,
the position of VLA1623B appears to move slightly with wavelength.
We return to this point below.

The total broadband continuum spectrum of all of VLA1623 was shown by AWB93 
(see also Andr\'e 1994). They fitted all of the data
shortward of 2~mm with a single temperature
modified blackbody, sometimes known as a
greybody. The monochromatic flux density, $S_\nu$, of a greybody, at 
frequency $\nu$, radiated into solid angle $\Omega$, is given by 

\begin{equation}
S_\nu = \Omega f B_\nu(T) [1 - e^{-(\nu/\nu_c)^\beta}] \, ,
\end{equation}

\noindent
where $B_\nu(T)$ is the blackbody function,
$\nu_c$ is the frequency at which the 
optical depth is unity, $\Omega$
is the solid angle of the aperture, $f$ is 
the filling factor of the source within the aperture and
$\beta$ is the dust emissivity index. 
The spectral index, $\alpha$, of a function on such a curve
is given by

\begin{equation}
\alpha = \frac{log(S_1/S_2)}{log(\nu_1/\nu_2)} \, ,
\end{equation}

\noindent
where $S_1$ and $S_2$ are the flux densities at frequencies $\nu_1$ and
$\nu_2$ respectively. In the long wavelength limit,
for a given temperature, $T$, the blackbody
function tends to

\begin{equation}
B_\nu \propto \nu^2 \, ,
\end{equation}

\noindent
which is known as the Rayleigh-Jeans limit. This is true in the
mm wavelength regime. In addition, if the emission is also optically
thin ($\nu << \nu_c$),
which it usually is for cold dust in the mm regime, then

\begin{equation}
S_\nu \propto \nu^{2+\beta} \, ,
\end{equation}

\noindent
from which we see that
$\beta$ is related to the
spectral index $\alpha$ in the mm by the equation

\begin{equation}
\alpha = 2 + \beta \, .
\end{equation}

\noindent
Hence a single-temperature, optically-thin greybody will produce a
straight line on a plot of log($S_\nu$) versus log($\nu$)
in the mm regime.
AWB93 and Andr\'e (1994) found temperatures of
15--20~K, with $\beta$=1.5$\pm$0.3 for the whole of VLA1623.

\begin{table}
\caption{Data for the two components of VLA1623. The flux densities are
quoted in mJy and measured in aperture sizes of 0.95$\times$0.4 arcsec,
1.0$\times$0.4 arcsec and 1.0$\times$0.75 arcsec at 2.7~mm, 7~mm and
3.6~cm respectively. The upper limits at 4.4~cm are measured in an 
interferometer beam size of 0.2$\times$0.05 arcsec, and 
hence the interferometer may have resolved away much of the flux density.
All flux densities are quoted to 2~s.f. and all errors to 1~s.f.
The 2.7-mm data are from Looney et al. (2000), and the 3.6-cm data are from
Bontemps \& Andr\'e (1997). The spectral index at 2.7~mm 
$\alpha_{mm}$ is taken to be 3.5, as for the whole of VLA1623 (AWB93).
This was subtracted from the longer wavelength
flux densities, and a spectral index at cm 
wavelengths $\alpha_{cm}$ was calculated for the residual flux densities.
The masses were simply estimated by scaling the 2.7-mm flux densities to
the total VLA1623 flux density (see text for discussion).
}
\begin{center}
\begin{tabular}{ccc}
\hline
Source & VLA1623A & VLA1623B \\
%& &  \\ 
\hline
%& &  \\
S$_{2.7mm}$ (mJy) & 22 $\pm$ 4      & 26 $\pm$ 4       \\
S$_{7mm}$ (mJy)   & 1.2 $\pm$ 0.2   & 2.0 $\pm$ 0.2    \\
S$_{3.6cm}$ (mJy) & 0.13 $\pm$ 0.02 & 0.40 $\pm$ 0.02  \\
S$_{4.4cm}$ (mJy) & $\leq$0.24      & $\leq$0.24       \\
$\alpha_{mm}$     & 3.5 $\pm$ 0.3   & 3.5 $\pm$ 0.3    \\
$\alpha_{cm}$     & 0.6 $\pm$ 0.3  & 0.6 $\pm$ 0.3   \\
Mass (M$_\odot$)  & $\leq$0.13      & $\leq$0.15        \\
%& &  \\
\hline
\end{tabular}
\end{center}
\end{table}

If we assume that the mm integrated flux density is optically thin, then
the mass, M, 
of the gas and dust can be calculated using the following equation

\begin{equation}
M = \frac{S_\nu D^2}{\kappa_\nu B_\nu} \, ,
\end{equation}

\noindent
where D is the distance to the source and $\kappa_\nu$ is the mass opacity 
of the gas (c.f. Andr\'e \& Montmerle 1994;
Kirk, Ward-Thompson \& Andr\'e 2005). We typically use a
value of $\kappa_{\rm 1.3mm}$ = 0.01 cm$^2$g$^{-1}$ (see e.g. Simpson et al.
2008 and references therein).
Using this method, AWB93 found a mass of 0.6~M$_\odot$ within a 2000~AU
diameter aperture around VLA1623.

\subsection{VLA1623A}

We can place an upper limit on the size of the
emission from the protostar VLA1623A of 0.7~arcsec, or 100~AU,
at an assumed distance of 139~pc (Mamajek 2008;
Simpson, Nutter \& Ward-Thompson 2008),
based on the 7-mm data presented in Figure~1.
The projected distance between VLA1623A and B is 1.2~arcsec,
corresponding to a projected separation of 170~AU. 

Pudritz et al. (1996) discovered a compact mm/submm source at the position
of the protostar VLA1623A at 0.85 and 1.3~mm, using the 
James Clerk Maxwell Telescope (JCMT) and 
Caltech Submillimetre Observatory (CSO) as a two-dish
interferometer. This source was unresolved in their observations.
They were able to place an upper limit on the diameter of this
source of $\sim$0.8~arcsec. They derived a different actual size
because they assumed a different distance.
However, Looney et al. (2000) claimed that the data could not rule out
VLA1623 being a binary source, due to the sparse u-v plane coverage of
the JCMT-CSO interferometer.

\begin{figure}
\includegraphics[angle=0,width=95mm]{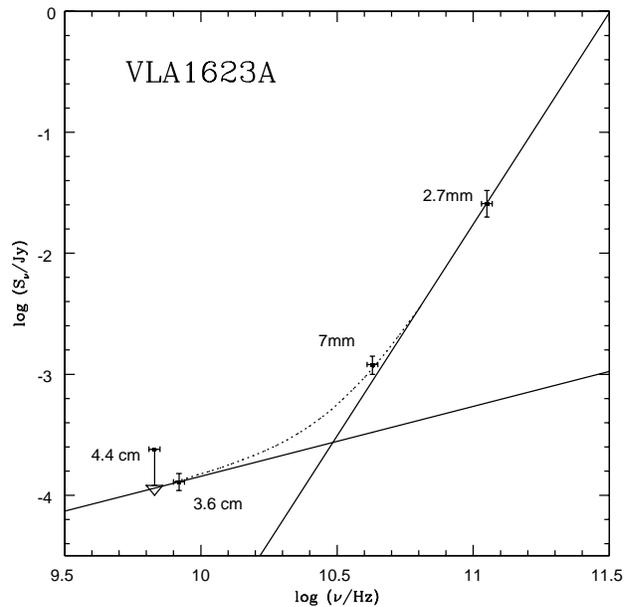}
\caption{Broadband spectrum of the source VLA1623A only, plotted as log
flux density, $S_\nu$, versus log frequency, $\nu$. Flux densities were
measured in each case on the compact source (see Table~1). 
This is speculated to be a circumstellar disc of radius $\sim$50~AU.
Note that a single straight line cannot fit the data. 
The steeper of the two solid lines has $\alpha$=3.5. and
$\beta$=1.5, consistent with the fit found for the whole
of VLA1623 (AWB93), normalised to the 2.7-mm flux density.
The flatter of the two solid lines has $\alpha$=0.6.
The dotted line is the sum of the two.
Data are taken from: 7~mm \& 4.4~cm -- this paper; 
3.6~cm -- Bontemps \& Andr\'e (1997); 2.7~mm -- Looney et al., (2000).}
\label{fig2}
\end{figure}

We postulate that the source we see at the position of the
VLA1623A protostar, with an upper limit diameter of 0.7~arcsec, is 
a circumstellar disc around VLA1623A. We do not resolve this source.
However, we note that the upper limit
radius of this suggested disc is 50~AU, consistent with
the sizes of discs around other young stellar objects
(e.g. Qi et al., 2004). The outflow from
VLA1623A is almost exactly in the plane of the sky 
(Andr\'e et al., 1990; Dent et al., 1995), so if this source
is a disc, we are probably seeing it close to edge-on.

Figure~2 shows the broad-band
spectrum of the compact VLA1623A protostar source on
its own, as a plot of log flux density against log frequency. The
data-points from Bontemps \& Andr\'e (1997), 
and Looney et al., (2000) are plotted on the same graph
as the flux density measured for the 
eastern source in Figure~1 (see also Table~1).
We did not detect VLA1623A at 4.4~cm with MERLIN, with a
1-$\sigma$ error-bar on the data
of $\sim$80~$\mu$Jy/beam at this wavelength. This is shown as a 3-$\sigma$
upper limit on Figure~2. 

However, we see that a single straight line does not fit the data.
Therefore, even on scales of 50~AU radius around the protostar
VLA1623A itself, there is more than one emission mechanism at work. 
This is to be expected. The disc
will contribute thermal dust emission. Similarly, there is predicted
to be a shock at the protostellar surface, which should produce
free-free emission.
Alternatively, there could be optically thick free-free emission from 
the base of the shock-ionized jet.

We cannot constrain the short wavelength spectral index, so we assume that
it has the same value as that for the whole of VLA1623, namely $\alpha$=3.5
and $\beta$=1.5 (AWB93). We acknowledge that this assumption may not be
valid, but it is the best assumption that can be made with the available data.
A grey-body function with these values
is shown as the steeper of the two solid lines 
on Figure~2, where it has simply been normalised to fit the 2.7-mm data.

The extrapolation of this curve was subtracted from the 7-mm and 3.6-cm
data, and the residual flux densities were fitted with a straight-line
fit. We refer to these residual flux densities as the dust-corrected 
flux densities. The fitted straight-line
is shown as the shallower of the two solid lines on Figure~2.
This line has a spectral index $\alpha$$\sim$0.6.
The dotted line shows the sum of the two solid lines. The dust spectrum 
only contributes a very small percentage to the 3.6-cm flux density.
Similarly, the shallow line only contributes a negligible percentage to
the 2.7-mm flux density. We note that the 7-mm data-point lies close to 
the break in the spectrum, and hence is crucial to its understanding.
The whole of VLA1623 also has a spectral break at around this point 
(Andr\'e 1994). The 4.4-cm upper limit lies above the fit to the cm
spectrum, so cannot constrain the spectral index.

If the mass of VLA1623A simply scaled with its 2.7-mm flux density
relative to the mass of the whole of VLA1623,
then the circumstellar disc around VLA1623A would have a mass of
$\leq$0.13~M$_\odot$. However, note that this assumes that the
temperature of the emitting dust remains the same from scales of 2000~AU
down to scales of $\sim$100~AU. It will probably be warmer, and
hence the mass will be lower. However, this also makes the assumption that 
the dust mass opacity remains the same across these different size-scales.
This would not be true if there were significant grain growth in the disc.
Pudritz et al. (1996) found
a lower limit to the disc mass of VLA1623A of $\sim$0.03~M$_\odot$,
and Looney et al. (2000) derive a lower limit of $\sim$0.04~M$_\odot$,
consistent with our results.

\begin{figure}
\includegraphics[angle=0,width=95mm]{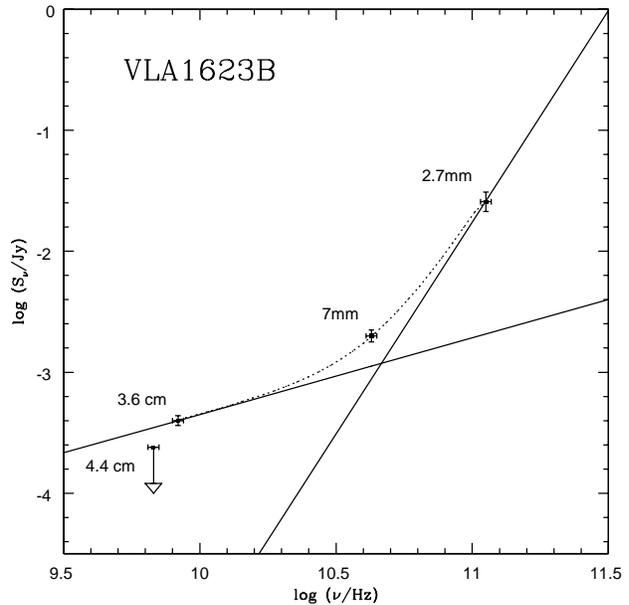}
\caption{Broadband spectrum of the source VLA1623B only, plotted as log
flux density, $S_\nu$, versus log frequency, $\nu$. Flux densities were
measured in each case on the western source
(see Table~1). Note how once again a single
straight line cannot fit the data. 
The steeper of the two solid lines has $\alpha$=3.5. and
$\beta$=1.5, consistent with the fit found for the whole
of VLA1623 (AWB93), normalised to the 2.7-mm flux density.
The flatter of the two solid lines has $\alpha$=0.6.
The dotted line is the sum of the two.
Data are taken from: 7~mm -- this paper; 
3.6~cm -- Bontemps \& Andr\'e (1997);
2.7~mm -- Looney et al., (2000).}
\label{fig3}
\end{figure}

The best-fit straight line to the 7-mm and 3.6-cm data has $\alpha$=0.6. 
This is greater than the spectral index of optically thin
free-free emission, which should be closer to $-$0.1. 
It could be that the free-free emission is optically thick,
and could arise at the base of the jet (Wright \& Barlow 1975; Reynolds 1986).
Alternatively, this excess could be due to a population of
larger grains, or planetesimals, growing in the disc around the protostar
VLA1623A (c.f. Greaves et al. 2008).

\subsection{VLA1623B}

Figure~3 shows the broad-band continuum
spectrum of the compact western source VLA1623B,
as a plot of log flux density against log frequency. The
data-points from Bontemps \& Andr\'e (1997) 
and Looney et al. (2000) are plotted on the same graph
as the flux density measured for this
source in Figure~1 (see also Table~1). 

Again we see that a single straight line does not fit the data.
So again there must be more than one emission mechanism at work. 
We cannot constrain the short wavelength spectral index, so we assume that
it has the same value as that for the whole of VLA1623, namely $\alpha$=3.5
and $\beta$=1.5 (AWB93), as before.
A grey-body function with these values
is shown as the steeper of the two solid lines 
on Figure~3, where it has been normalised to fit the 2.7-mm data.

The extrapolation of this curve was subtracted from the 7-mm and 3.6-cm
data, and the residual dust-corrected
flux densities were fitted with a straight-line
fit, as before. The fitted straight-line
is shown as the shallower of the two solid lines on Figure~3.
This line has a spectral index $\alpha$=0.6.
The dotted line shows the sum of the two solid lines. The dust spectrum 
only contributes a very small percentage to the 3.6-cm flux density.
Similarly, the shallow line only contributes a negligible percentage to
the 2.7-mm flux density.

If the mass of VLA1623B simply scaled with its 2.7-mm flux density
relative to the mass of the whole of VLA1623,
then VLA1623B would have a mass of
$\leq$0.15~M$_\odot$.
Once again we note that this assumes that the
temperature of the emitting dust remains the same from scales of 2000~AU
down to scales of $\sim$100~AU. It also makes the assumption that the
dust mass opacity remains the same across these different size-scales.

We did not detect VLA1623B at 4.4~cm with MERLIN, with a
1-$\sigma$ error-bar on the data
of $\sim$80~$\mu$Jy/beam at this wavelength. This is shown as a 3-$\sigma$
upper limit on Figure~3. We see that this lies below the extrapolated
fit to the cm flux densities. Hence,
VLA1623B cannot be a compact source relative to our beam.
This would suggest that the source is extended
relative to our beam of $\sim$0.2 $\times$ 0.05 arcsec, which is approximately
$\sim$30 $\times$ 7~AU. This puts a lower limit to the size of VLA1623B.

\section{Discussion}

The nature of VLA1623B remains a matter for debate. It is either a binary
protostar component, 
as suggested by Looney et al. (2000), or it is a knot in the
radio jet, as suggested by Bontemps \& Andr\'e (1997). We now summarise 
the evidence for and against each of these hypotheses.

\subsection{Binary protostar?}

The case for VLA1623 being a binary protostar appears strong when one looks
at the 2.7-mm data (Looney et al. 2000).
The sources appear to be of similar size and have
similar flux densities at this wavelength. 
Likewise, the similarity of their cm spectral indices appears
to indicate that these two sources are the same type of object.

However, the nature of such a 
binary must be somewhat more complex than our simple ideas would predict.
For example, one might normally expect the two components of a binary system
to have similar angular momenta.
In this case the outflow from VLA1623A is directed roughly towards VLA1623B.
Hence the presumed disc around VLA1623A cannot be co-planar with the line
joining VLA1623A \& B, which would be the plane of their mutual orbits.
Therefore, the angular momentum of the VLA1623A disc would be misaligned
relative to the angular momentum of the putative binary protostar.

This does not preclude this hypothesis. Other misaligned jets in binary
systems have been observed (e.g. Albrecht et al. 2009).
They have also been modelled (e.g. Stamatellos et al., 2011; Walsh et al.,
in prep).
However, it does mean that it is not a simple binary protostellar system.

\subsection{Knot in the jet?}

The case for VLA1623B being a knot in the radio jet from VLA1623A appears
strong when one looks at the 3.6-cm radio data (Bontemps \& Andr\'e 1997).
There are several radio sources in a line apparently emanating from
VLA1623A. At radio wavelengths VLA1623B is also much brighter than VLA1623A.
This would tend to lend support to the hypothesis that the emission is
arising from shocks in the jet.

Furthermore, close examination of Figure~1 shows that, whilst the position of
VLA1623A appears constant with wavelength, that of VLA1623B does not.
The peak of the 7-mm data is offset from that of the 3.6-cm data, and both
are offset from that of the 2.7-mm data. This would lend support to the
hypothesis that this is not a single-peaked protostellar source.

However, the putative radio jet is not exactly coincident with the CO
outflow (Andr\'e et al. 1990; Dent et al. 1995), but rather is at an angle
to it. Nevertheless, this may not be a problem. It could be that the
3.6-cm radio sources (including VLA1623B) could be the positions where a
radio jet is interacting with the wall of the cavity caused by the CO
outflow. In this hypothesis the heated dust emission seen at 2.7~mm would
be shock-heated dust on the cavity wall.

Similar sources have been seen elsewhere by Maury et al. (2010). They
see two apparent components to the protostar NGC1333-IRAS2A. They see
something similar in L1448C. In this case the two components lie along a
similar direction to the axis of the outflow (see their figure~3), but
offset by a small angle, exactly as we see in VLA1623. They interpret
the secondary source as a shock-heated portion of the outflow cavity wall.

\section{Conclusions}

We have presented mm and cm interferometer data of the Class~0 protostar
VLA1623. We have detected the two components A \& B that were previously 
known. The projected separation of the two components is $\sim$170~AU.
VLA1623A is a protostar with a circumstellar disc,
whose radius lies in the range of $\sim$10--50~AU. We estimate a
limit to the disc mass of $\leq$0.13~M$_\odot$.
Excess emission at cm wavelengths is most likely due to
optically thick free-free emission,
but it could be indicating the onset of
grain growth or planetesimal formation in the disc.
The nature of VLA1623B remains a matter for debate. It lies along the line
of the radio jet, and may be a shocked knot in the radio jet.
Alternatively, it may be a binary component
of the protostar. Further data are required to resolve this question.

\section*{Acknowledgements}

MERLIN is a UK national facility operated by the University of Manchester 
at Jodrell Bank Observatory (JBO) on behalf of the UK Science and
Technology Facilities Council (STFC).
This work has benefited from research funding from the European 
Community's sixth Framework Programme under
RadioNet R113CT 2003 5058187.
The authors would like to thank the staff of the 
JBO for assistance during the taking and reducing of the MERLIN data.
In particular, we thank Anita Richards and Tom Muxlow for their time and
patience during the rather tricky data reduction phase.
The VLA is operated by the American National Radio Astronomy 
Observatory (NRAO). NRAO is a facility of the American
National Science Foundation (NSF), operated under
co-operative agreement by Associated Universities, Inc.
JMK acknowledges STFC for post-doctoral
support through the Cardiff Astronomy Rolling Grant. 
The authors wish to thank the anonymous referee for helpful
comments on an earlier draft of this paper.

\end{document}